\documentclass[preprint,showpacs,preprintnumbers,amsmath,amssymb,endfloats*]{revtex4}
\usepackage{graphicx}

\newcommand{\bes}{\begin{eqnarray}}
\newcommand{\ees}{\end{eqnarray}}

\begin{document}

\thispagestyle{empty}
\title{Van der Waals interaction between microparticle and
uniaxial crystal with application to hydrogen
atoms and multiwall carbon nanotubes
}
\author{E.~V.~Blagov,${}^{1}$
  G.~L.~Klimchitskaya,${}^{2,}$\footnotemark[1]
  and
V.~M.~Mostepanenko${}^{1,}$\footnote[1]{Present address: Institute for
Theoretical Physics, Leipzig University, Augustusplatz 10/11,
04109, Leipzig, Germany.}
}

\affiliation{${}^{1}$Noncommercial Partnership ``Scientific Instruments'', 
Tverskaya St. 11, Moscow, 103905, Russia \\
${}^{2}$North-West Technical University, Millionnaya St. 5, St.Petersburg,
191065, Russia
}

\begin{abstract}
The Lifshitz theory of the van der Waals force is extended for
the case of an atom (molecule) interacting with a plane surface
of an uniaxial crystal or with a long solid cylinder or
cylindrical shell made of isotropic material or uniaxial crystal.
For a microparticle near a semispace or flat plate made of an
uniaxial crystal the exact expressions for the free energy of
the van der Waals and Casimir-Polder interaction are presented.
 An approximate
expression for the free energy of microparticle-cylinder interaction
is obtained which becomes precise
for microparticle-cylinder separations much smaller than cylinder
radius. The obtained expressions are used to investigate the
van der Waals interaction between hydrogen atoms (molecules) and
graphite plates or multiwall carbon nanotubes. To accomplish this
the behavior of graphite dielectric permittivities along the
imaginary frequency axis is found using the optical data for
the complex refractive index of graphite for the ordinary and extraordinary 
rays. It is shown that the position of hydrogen atoms inside 
multiwall carbon nanotubes is energetically preferable compared
with outside.
\end{abstract}


\maketitle

\section{Introduction}

The van der Waals interaction between microparticle and macrobody
has long been investigated. It is of much importance for
understanding of a large body of physical and chemical phenomena
connected with atom-surface interaction including adsorption and
friction. In a pioneering work in Ref.~\cite{1}, the interaction potential
between an atom at a separation $a$ from a plane wall was found
in the from $V_3(a)=-C_3/a^3$. This result is applicable at
separations less than a few nanometers.
More recently, a lot of different atoms, molecules and wall materials
was studied. In particular, in Refs.~\cite{1a,1b} the values of
$C_3$ were computed for the interaction of H, H${}_2$, He, Ne, Ar,
Cr, Xe, and CH${}_4$ with the planar surfaces of insulators (sapphire, LiF,
CaF${}_2$, and boron nitride).
 At much greater separations the atom-wall interaction
is described by the Casimir-Polder potential $V_4(a)=-C_4/a^4$
\cite{2} taking relativistic effects into account.
The complete theory of the van der Waals atom-wall interaction
at nonzero temperature is given by the Lifshitz formula \cite{3}
in terms of the dynamic polarizability of an atom (molecule)
and the frequency-dependent dielectric permittivity of wall
material. The potentials $V_3(a)$ and $V_4(a)$, obtained
previously, are the two limiting cases of this formula.

During the last few years van der Waals forces have found important 
new applications in experiments on quantum reflection and 
diffraction of ultra-cold atoms on different surfaces \cite{4,5,6,7}
and in Bose-Einstein condensation \cite{8,9}. In connection with
this, the detailed examination of different corrections to the
Casimir-Polder and van der Waals interactions, including 
the precise effect of atomic
polarizability and nonideality of wall material was performed
in Refs.~\cite{10,10a}. Effectively this resulted in the investigation
of accurate dependences of the coefficients $C_3$ and $C_4$ on
separation and temperature.

Although the Lifshitz theory presents considerable opportunity
for extensive studies of the van der Waals force \cite{11,12},
it is essentially restricted by macroscopic bodies with plane
boundaries. The use of approximations, like the proximity force
theorem \cite{13}, permitted one to obtain rather precise results
for a large sphere near a plane plate, a configuration frequently
used in recent experiments on measuring the Casimir force
\cite{14,15,16,17,18}. In most cases the macrobodies with
plane boundaries were supposed to be isotropic.

In the present paper we generalize the Lifshitz formula for a
microparticle situated near the surface of an uniaxial crystal.
Both cases of crystal semispace with plane boundary and a plane
plate of finite thickness are considered. As a next step, we
derive the approximate expression for the free energy of the 
van der Waals interaction between a microparticle and a solid
cylinder or cylindrical shell made of an uniaxial crystal. 
In the limiting case this expression is applicable to a
microparticle near a cylinder made of an isotropic material with
frequency dependent dielectric permittivity (a configuration
which also has not been investigated previously).
We apply the
obtained results to investigate the van der Waals interaction
between hydrogen atoms or molecules and graphite plates or
multiwall carbon nanotubes.

The study of the van der Waals interaction between hydrogen atoms 
and a graphitic surface has become urgent after the proposal of
Ref.~\cite{19} to use the singlewall carbon nanotubes for
hydrogen storage. Since, many papers were published on the use of
both singlewall and multiwall nanotubes for hydrogen storage and
containing both promising and disappointing results (see
Ref.~\cite{20} for review). The macroscopic theoretical approach
leads to a conclusion \cite{21} that the carbon nanostructures
might absorb hydrogen from 4 to 14 percent of
their weight. However, the microscopic mechanisms responsible
for this absorption are still unknown. The van der Waals forces
acting between hydrogen atoms or molecules and carbon nanostructures,
which might play an important role in absorption phenomena, are
practically unexplored. Some preliminary results for graphite sheets
and singlewall nanotubes can be found in Refs.~\cite{22} and 
\cite{23,23a}, respectively. The van der Waals interaction of
fulerene molecules and adsorption of these molecules on graphite
were considered in Ref.~\cite{23b}.

To apply the Lifshitz-type formulas for the van der Waals free
energy, obtained in the paper, to the case of hydrogen atoms and
molecules near graphite surface, we calculate the dielectric
permittivities of graphite and dynamic polarizabilities of hydrogen
atom and molecule along the imaginary frequency axis. To do this,
we discuss different sets of tabulated optical data for the complex
refractive index of graphite and use the most reliable ones to perform
the Kramers-Kronig analysis. The van der Waals interactions between
hydrogen atom and molecule and graphite semispace or plate of finite
thickness are calculated. The free energies of hydrogen atom inside and 
outside of a multiwall carbon nanotube are found as functions of
an atom-nanotube separation distance and internal and external nanotube
radia. The location of a hydrogen atom inside a multiwall nanotube is
demonstrated to be preferable from an energetic point of view.

The paper is organized as follows. In Sec.~II we present the Lifshitz
formula for the van der Waals (and Casimir-Polder) interaction
between microparticle and plane surface of an uniaxial crystal.
Sec.~III contains derivation of general expression for the van der
Waals free energy of a microparticle external to a solid cylinder or
cylindrical shell
made of an uniaxial crystal. In Sec.~IV the dielectric permittivities 
of graphite and the atomic and molecular dynamic
polarizabilities of hydrogen along the imaginary
frequency axis are obtained. In Sec.~V calculation results are presented
for the van der Waals interaction between
hydrogen atom or molecule and graphite semispace or a plane plate
of finite thickness. In Sec.~VI the same is done for   
hydrogen atom or molecule external to a multiwall carbon nanotube.
Comparison between the free energies of
hydrogen atom inside and outside multiwall nanotube is done in
Sec.~VII. Sec.~VIII contains our discussion and conclusions.

\section{Lifshitz formula for the van der Waals interaction 
\protect{\\}
between
microparticle and plane surface of an uniaxial crystal}

First we consider a neutral microparticle (atom or molecule) with
a dynamic polarizability $\alpha(\omega)$ at separation $a$ from
a plane surface of the isotropic semispace with dielectric
permittivity $\varepsilon(\omega)$ at temperature $T$ in thermal
equilibrium. In this case the free energy of 
microparticle-semispace van der Waals interaction is given by the
familiar Lifshitz formula \cite{3} (see also \cite{9,10,24,25,26})
\begin{eqnarray}
&&
F_E^s(a,T)=-k_BT\sum\limits_{l=0}^{\infty}
{\vphantom{\sum}}^{\prime}\alpha(i\xi_l)
\int_{0}^{\infty}k_{\bot}dk_{\bot}
q_le^{-2aq_l}\nonumber \\
&&\phantom{aaa}
\times\left\{2r_{\|}^{s}(\xi_l,k_{\bot})+
\frac{\xi_l^2}{q_l^2c^2}\left[
r_{\bot}^{s}(\xi_l,k_{\bot})-r_{\|}^{s}(\xi_l,k_{\bot})
\right]\right\}.
\label{e1}
\end{eqnarray}
\noindent
Here $\xi_l=2\pi k_BTl/\hbar$ are the Matsubara frequencies,
$k_B$ is the Boltzmann constant, $l=0,\,1,\,2,\,\ldots\,$,
and $k_{\bot}$ is the magnitude of a wave vector component in
the plane surface of a semispace. The coefficients of reflection
for two independent polarizations of electromagnetic field are
given by
\begin{eqnarray}
&&r_{\|}^s(\xi_l,k_{\bot})=
\frac{\varepsilon_lq_l-k_l}{\varepsilon_lq_l+k_l},
\nonumber \\
&&r_{\bot}^s(\xi_l,k_{\bot})=\frac{k_l-q_l}{k_l+q_l},
\label{e2}
\end{eqnarray}
\noindent
where
\begin{eqnarray}
&&
q_l=\sqrt{k_{\bot}^2+\frac{\xi_l^2}{c^2}},\quad
k_l=\sqrt{k_{\bot}^2+\varepsilon_l\frac{\xi_l^2}{c^2}},
\nonumber \\ &&
\varepsilon_l=\varepsilon(i\xi_l),
\label{e3}
\end{eqnarray}
\noindent
and prime near the summation sign in Eq.~(\ref{e1}) means that the
term for $l=0$ has to be multiplied by 1/2.

Eq.~(\ref{e1}) can be readily generalized for the case when the
microparticle is located not near a semispace, but near a flat
plate of some finite thickness $d$ with the same dielectric
permittivity $\varepsilon(\omega)$. In this case the free energy
of the van der Waals interaction $F_E^p(a,T)$ again is given
by Eq.~(\ref{e1})  where, however, the reflection coefficients
from a semispace $r_{\|,\bot}^{s}(\xi_l,k_{\bot})$ should be
replaced by the reflection coefficients from a plate of
finite thickness $r_{\|,\bot}^{p}(\xi_l,k_{\bot})$.
The explicit expressions for them are obtained from the free
energy of the van der Waals interaction between the layered
media (see, e.g., \cite{24,27,28}):
\begin{eqnarray}
&&r_{\|}^p(\xi_l,k_{\bot})=
\frac{\varepsilon_l^2q_l^2-k_l^2}{\varepsilon_l^2q_l^2+
k_l^2+2q_lk_l\varepsilon_l\mbox{coth}(k_ld)},
\nonumber \\
&&r_{\bot}^p(\xi_l,k_{\bot})=\frac{k_l^2-q_l^2}{k_l^2+
q_l^2+ 2q_lk_l\mbox{coth}(k_ld)}.
\label{e4}
\end{eqnarray}
\noindent
In the limit $d\to\infty$ Eq.~(\ref{e4}) transforms into Eq.~(\ref{e2}).

Let us now consider a semispace or a plate of finite thickness made
of an uniaxial crystal (graphite for instance) which is characterized
by two dissimilar dielectric permittivities
$\varepsilon_x(\omega)=\varepsilon_y(\omega)$ and
$\varepsilon_z(\omega)$. Let a microparticle be located near the
uniaxial crystal semispace restricted by the plane $(x,y)$, and the 
crystal optical axis $z$ being perpendicular to it. Then the
free energy of the van der Waals interaction is again given by
Eq.~(\ref{e1}) where the coefficients of reflection from
the surface of isotropic semispace $r_{\|,\bot}^s(\xi_l,k_{\bot})$
should be replaced by their generalization for the case of
uniaxial crystal (graphite) \cite{29}:
\begin{eqnarray}
&&r_{\|;g}^s(\xi_l,k_{\bot})=
\frac{\sqrt{\varepsilon_{xl}\varepsilon_{zl}}q_l-
k_{zl}}{\sqrt{\varepsilon_{xl}\varepsilon_{zl}}q_l+k_{zl}},
\nonumber \\
&&r_{\bot;g}^s(\xi_l,k_{\bot})=
\frac{k_{xl}-q_l}{k_{xl}+q_l}.
\label{e5}
\end{eqnarray}
\noindent
Here the following notations are introduced
\begin{eqnarray}
&&k_{xl}=\sqrt{k_{\bot}^2+\varepsilon_{xl}
\frac{\xi_l^2}{c^2}}, \qquad
k_{zl}=\sqrt{k_{\bot}^2+\varepsilon_{zl}
\frac{\xi_l^2}{c^2}},
\nonumber \\
&&\varepsilon_{xl}=\varepsilon_x(i\xi_l), \qquad
\varepsilon_{zl}=\varepsilon_z(i\xi_l).
\label{e6}
\end{eqnarray}
\noindent
For isotropic crystal $\varepsilon_x=\varepsilon_z=\varepsilon$
and Eq.~(\ref{e5}) coincides with  Eq.~(\ref{e2}).

If a microparticle is located near a flat plate of finite thickness
made of uniaxial crystal ($z$-axis is perpendicular to the plate),
the free energy $F_E^p(a,T)$ is given again by Eq.~(\ref{e1}),
where the coefficients of reflection from an isotropic plate
$r_{\|,\bot}^p(\xi_l,k_{\bot})$ are replaced by the reflection
coefficients from a plate made of uniaxial crystal:
\begin{eqnarray}
&&r_{\|;g}^p(\xi_l,k_{\bot})=
\frac{{\varepsilon_{xl}\varepsilon_{zl}}q_l^2-
k_{zl}^2}{{\varepsilon_{xl}\varepsilon_{zl}}q_l^2+k_{zl}^2+
2\sqrt{\varepsilon_{xl}\varepsilon_{zl}}q_lk_{zl}
\mbox{coth}(k_{zl}d)},
\nonumber \\
&&r_{\bot;g}^p(\xi_l,k_{\bot})=
\frac{k_{xl}^2-q_l^2}{k_{xl}^2+q_l^2+
2q_lk_{xl}\mbox{coth}(k_{xl}d)}.
\label{e7}
\end{eqnarray}
\noindent
For the anisotropic plate of infinite thickness ($d\to\infty$)
Eq.~(\ref{e7}) transforms into Eq.~(\ref{e5}). On the other hand,
in the limit of the plate made of isotropic substance Eq.~(\ref{e7})
coincides with Eq.~(\ref{e4}).

Eq.~(\ref{e1}) with reflection coefficients (\ref{e5}), (\ref{e7})
is used in Sec.V for computations of the van der Waals interaction
between the hydrogen atoms or molecules and the plane surface of
a semispace or a plate made of graphite.

\section{Free energy of the van der Waals interaction for
\protect{\\}
a microparticle external to a solid or hollow cylinder}

In this section we derive the Lifshitz-type formula for the van der
Waals free energy of a microparticle located at a separation $a$
from the external surface of a solid cylinder or cylindrical shell made
of an uniaxial crystal. It is assumed that the crystal optical
axis $z$ is perpendicular to the cylinder surface of crystalline
layers. The outer radius of a cylinder is $R$ and the thickness of
a crystal cylindrical shell is $d\leq R$. In the case $d=R$
the cylinder is solid. If $d<R$, there is an empty cylindrical cavity
inside of a cylinder. As in the previous section, the crystalline material
of the cylindrical shell is described by the dielectric permittivities
$\varepsilon_x(\omega)$ and $\varepsilon_z(\omega)$. The derivation
presented below is based on the same approach which was previously
used in literature \cite{3,10,24,25,26} to derive the Lifshitz
formula for microparticle-semispace (plate) interaction from the 
Lifshitz formula for a configuration of two parallel semispaces (plates).

Let us consider an infinite space filled with an isotropic
substance having a dielectric permittivity $\varepsilon(\omega)$,
containing an empty cylindrical cavity of radius $R+a$. We introduce our
solid cylinder or cylindrical shell of external radius $R$ made of an 
uniaxial crystal inside this cavity so that the cylinder axis coincides
with the axis of the cavity (see Fig.~1). Then there is a gap of
thickness $a$ between our cylinder and the boundary of the cylindrical 
cavity of radius $R+a$ restricting the infinite space with the
dielectric permittivity $\varepsilon(\omega)$. Each element of our
cylinder experiences an attractive van der Waals interaction
on the source side of the boundary of the cylindrical cavity restricting
the infinite space. With the help of the proximity force theorem
the free energy of this interaction between two cylinders can be
approximately represented in the form (see Ref.~\cite{30} for
the case of ideal metals)
\begin{equation}
F_E^{c,c}(a,T)=2\pi L\sqrt{R(R+a)}F_E^{i,s}(a,T).
\label{e8}
\end{equation}
\noindent
Here $F_E^{i,s}(a,T)$ is the free energy per unit area in the
configuration either of two semispaces separated by a gap of
width $a$ (in this case $i=s$, our cylinder is solid, one
semispace is filled with an uniaxial crystal and the other is
filled with a material of dielectric permittivity 
$\varepsilon(\omega)$) or of a flat plate of thickness $d$ and
a semispace separated by the same gap (in this case $i=p$, 
and we are dealing with cylindrical shell having a longitudinal hole of
radius $R-d$; the plate is made of an uniaxial crystal and 
semispace of material with a dielectric permittivity 
$\varepsilon(\omega)$). In Eq.~(\ref{e8}) $L$ is the length of
our solid or hollow cylinder which is supposed to be much larger
than its radius $R$.

As shown in Ref.~\cite{30} (see also Ref.~\cite{31}), the accuracy
of Eq.~(\ref{e8}) is rather high. For example, within the separation
region $0<a<R/2$ the results calculated by Eq.~(\ref{e8}) coincide
with the exact ones up to 1\% in the case of cylinders made of
perfect metal (for other materials the accuracy may be different for
only a fraction of percent). This is quite satisfactory for
application to multiwall nanotubes with $R$ of about a few ten
nanometers considered below.

The explicit expressions for the free energy $F_E^{i,s}(a,T)$
are well known \cite{3,24,25,26,27,28}
\begin{eqnarray}
&&
F_E^{i,s}(a,T)=\frac{k_BT}{2\pi}
\sum\limits_{l=0}^{\infty}{\vphantom{\sum}}^{\prime}
\int_{0}^{\infty}k_{\bot}dk_{\bot}\left\{
\ln\left[1-r_{\|;g}^{s,p}(\xi_l,k_{\bot})
r_{\|}^{s}(\xi_l,k_{\bot})e^{-2aq_l}\right]\right.
\nonumber \\
&&\phantom{aaaaa}
+\left.\ln\left[1-r_{\bot;g}^{s,p}(\xi_l,k_{\bot})
r_{\bot}^{s}(\xi_l,k_{\bot})e^{-2aq_l}\right]\right\}.
\label{e9}
\end{eqnarray}
\noindent
Here the reflection coefficients
$r_{\|,\bot;g}^s$ 
from the semispace of uniaxial crystal are given by 
Eq.~(\ref{e5}),  coefficients $r_{\|,\bot;g}^p$, describing
reflection from a flat plate of uniaxial crystal, are given by
Eq.~(\ref{e7}), and coefficients $r_{\|,\bot}^s$ describing
reflection from isotropic semispace are presented in Eq.~(\ref{e2}).
Notice that when index $i$ in the left-hand side of Eq.~(\ref{e9})
is equal to $s$ or $p$ one should choose $s$ or $p$ in the
right-hand side, respectively.

To continue with our derivation, we now suppose that the isotropic
substance with the dielectric permittivity $\varepsilon(\omega)$
is rarefied with the number $N$ of atoms or molecules per unit
volume. Expanding the quantity $F_E^{c,c}(a,T)$ from the left-hand
side of Eq.~(\ref{e8}) as a power series in $N$ and using the
additivity of the first-order term, one can write
\begin{equation}
F_E^{c,c}(a,T)=N\int_{a}^{\infty}
F_E^{c}(z,T)2\pi (R+z)Ldz+O(N^2),
\label{e10}
\end{equation}
\noindent
where $F_E^{c}(z,T)$ is the free energy of the van der Waals
interaction of a single atom belonging to an isotropic substance
with a solid cylinder or cylindrical shell made of an uniaxial crystal
(note that separation $z$ is measured from the external surface
of the cylinder in the direction perpendicular to it).

By differentiation of both sides of Eq.~(\ref{e10}) with respect to
$a$, we obtain
\begin{equation}
-\frac{\partial F_E^{c,c}(a,T)}{\partial a}=
2\pi (R+a)LNF_E^{c}(a,T)+O(N^2).
\label{e11}
\end{equation}

The same derivative can be found when differentiating
both sides of Eq.~(\ref{e8})
\begin{eqnarray}
&&
-\frac{\partial F_E^{c,c}(a,T)}{\partial a}=
2\pi L\sqrt{R(R+a)}
\label{e12} \\
&&\phantom{aaaaaaaa}
\times\left[-\frac{1}{2(R+a)}F_E^{i,s}(a,T)
+ F^{i,s}(a,T)
\vphantom{\frac{1}{2(R+a)}}\right],
\nonumber
\end{eqnarray}
\noindent
where
\begin{equation}
F^{i,s}(a,T)=-\frac{\partial F_E^{i,s}(a,T)}{\partial a}
\label{e13}
\end{equation}
\noindent
is the van der Waals force per unit area acting between the semispace
made of an uniaxial crystal ($i=s$) or a flat plate made of the
same material and a semispace with a dielectric permittivity
$\varepsilon$. The expression for this force is easily obtained
from Eqs.~(\ref{e9}) and (\ref{e13}):
\begin{eqnarray}
&&
F^{i,s}(a,T)=-\frac{k_BT}{\pi}
\sum\limits_{l=0}^{\infty}{\vphantom{\sum}}^{\prime}
\int_{0}^{\infty}k_{\bot}dk_{\bot}q_l
\label{e14} \\
&&\phantom{aaaa}
\times\left[\frac{r_{\|;g}^{s,p}(\xi_l,k_{\bot})
r_{\|}^{s}(\xi_l,k_{\bot})}{e^{2aq_l}-
r_{\|;g}^{s,p}(\xi_l,k_{\bot})
r_{\|}^{s}(\xi_l,k_{\bot})}
+\frac{r_{\bot;g}^{s,p}(\xi_l,k_{\bot})
r_{\bot}^{s}(\xi_l,k_{\bot})}{e^{2aq_l}-
r_{\bot;g}^{s,p}(\xi_l,k_{\bot})
r_{\bot}^{s}(\xi_l,k_{\bot})}\right].
\nonumber
\end{eqnarray}

The dielectric permittivity of a rarefied substance can be
expanded in Taylor series in powers of $N$ \cite{32}
\begin{equation}
\varepsilon(i\xi_l)=1+4\pi\alpha(i\xi_l)N+O(N^2),
\label{e15}
\end{equation}
\noindent
where
$\alpha(\omega)$ is the dynamic polarizability of an atom
(molecule) of this substance. Substituting Eq.~(\ref{e15})
in Eqs.~(\ref{e2}) and (\ref{e3}) we obtain
\begin{eqnarray}
&&
r_{\|}^{s}(\xi_l,k_{\bot})=\pi\alpha(i\xi_l)N
\left(2-\frac{\xi_l^2}{q_l^2c^2}\right)+O(N^2),
\nonumber \\
&&
r_{\bot}^{s}(\xi_l,k_{\bot})=\pi\alpha(i\xi_l)
\frac{N\xi_l^2}{q_l^2c^2}+O(N^2).
\label{e16}
\end{eqnarray}

Using Eq.~(\ref{e16}), the free energy $F_E^{i,s}$ and the force
$F^{i,s}$ from Eqs.~(\ref{e9}) and (\ref{e14}) can be represented
in the form
\begin{eqnarray}
&&
F_E^{i,s}(a,T)=-\frac{k_BTN}{2}
\sum\limits_{l=0}^{\infty}{\vphantom{\sum}}^{\prime}
\alpha(i\xi_l)
\int_{0}^{\infty}k_{\bot}dk_{\bot}
\nonumber \\
&&\phantom{aaa}
\times
\left[\left(2-\frac{\xi_l^2}{q_l^2c^2}\right)
r_{\|;g}^{s,p}(\xi_l,k_{\bot})+
\frac{\xi_l^2}{q_l^2c^2}
r_{\bot;g}^{s,p}(\xi_l,k_{\bot})\right]
e^{-2aq_l}+O(N^2),
\nonumber \\
&& \label{e17} \\
&&
F^{i,s}(a,T)=-k_BTN
\sum\limits_{l=0}^{\infty}{\vphantom{\sum}}^{\prime}
\alpha(i\xi_l)
\int_{0}^{\infty}k_{\bot}dk_{\bot}q_l
\nonumber \\
&&\phantom{aaa}
\times
\left[\left(2-\frac{\xi_l^2}{q_l^2c^2}\right)
r_{\|;g}^{s,p}(\xi_l,k_{\bot})+
\frac{\xi_l^2}{q_l^2c^2}
r_{\bot;g}^{s,p}(\xi_l,k_{\bot})\right]
e^{-2aq_l}+O(N^2).
\nonumber 
\end{eqnarray}
\noindent

Substituting Eq.~(\ref{e17}) in Eq.~(\ref{e12}), one finds
\begin{eqnarray}
&&
-\frac{\partial F_E^{c,c}(a,T)}{\partial a}=-2\pi LNk_BT
\sqrt{R(R+a)}
\sum\limits_{l=0}^{\infty}{\vphantom{\sum}}^{\prime}
\alpha(i\xi_l)
\int_{0}^{\infty}k_{\bot}dk_{\bot}
\left[q_l-\frac{1}{4(R+a)}\right]
\nonumber \\
&&
\times
\left\{
2r_{\|;g}^{s,p}(\xi_l,k_{\bot})+
\frac{\xi_l^2}{q_l^2c^2}\left[
r_{\bot;g}^{s,p}(\xi_l,k_{\bot})-
r_{\|;g}^{s,p}(\xi_l,k_{\bot})\right]\right\}
e^{-2aq_l}+O(N^2).
\label{e18}
\end{eqnarray}

As a final stage of the derivation, we substitute the result (\ref{e18})
into the left-hand side of Eq.~(\ref{e11}), take the limit $N\to 0$ and 
arrive at desired expression for the free energy of van der Waals
interaction between a microparticle and a cylinder made of uniaxial
crystal
\begin{eqnarray}
&&
F_E^{c}(a,T)=-k_BT
\sqrt{\frac{R}{R+a}}
\sum\limits_{l=0}^{\infty}{\vphantom{\sum}}^{\prime}
\alpha(i\xi_l)
\int_{0}^{\infty}k_{\bot}dk_{\bot}e^{-2aq_l}
\left[q_l-\frac{1}{4(R+a)}\right]
\nonumber \\
&&\phantom{aaa}
\times
\left\{
2r_{\|;g}^{s,p}(\xi_l,k_{\bot})+
\frac{\xi_l^2}{q_l^2c^2}\left[
r_{\bot;g}^{s,p}(\xi_l,k_{\bot})-
r_{\|;g}^{s,p}(\xi_l,k_{\bot})\right]\right\}.
\label{e19}
\end{eqnarray}
\noindent
In the case of a solid cylinder, the reflection coefficients
$r_{\|,\bot;g}^s$, given by Eq.~(\ref{e5}), should be chosen in
the right-hand side of Eq.~(\ref{e19}). For a cylindrical shell,
coefficients $r_{\|,\bot;g}^p$ from Eq.~(\ref{e7}) should be
used. Notice that in the limit $R\to\infty$ Eq.~(\ref{e19})
coincides with a known result (\ref{e1}) for the free energy
of microparticle near a plane surface of a semispace.
The above derivation is preserved also in the limiting case of
a solid or hollow cylinder made of isotropic material with
$\varepsilon_x=\varepsilon_y=\varepsilon_z\equiv\varepsilon$.
To obtain the result for isotropic cylinder, one should substitute
in Eq.~(\ref{e19}) the reflection coefficients (\ref{e2}),
(\ref{e4}) instead of (\ref{e5}), (\ref{e7}).

Eq.~(\ref{e19}) is the approximate one. It is, however, practically
exact at $a\ll R$ and is of high precision (the error is of about
1\%) at all separations $a\leq R/2$. That is why this equation is
reliable for calculations of the van der Waals interaction between
a cylinder and microparticles located in its close proximity.

\section{Dielectric permittivities of graphite and dynamic
\protect{\\}
polarizabilities of hydrogen atom and molecule along the imaginary
frequency axis}

Below we used the Lifshitz-type formulas obtained above to
calculate the van der Waals interaction between hydrogen atoms or
molecules and graphite semispace or flat plate 
[Eqs.~(\ref{e1}),\ (\ref{e5}),\ (\ref{e7})] or graphite cylinder
[Eqs.~(\ref{e5}),\ (\ref{e7}),\ (\ref{e19})]. The graphite cylinder models
a multiwall carbon nanotube (see Sec.~VI). 
To attain these ends, one needs the
values of dynamic polarizabilities of hydrogen atom and molecule
and also both dielectric permittivities of graphite at all
Matsubara frequencies which give non-negligible contribution to
the result.

The precise expression for the atomic dynamic polarizability of
hydrogen is given by the 10-oscillator formula \cite{33}
written in atomic units
\begin{equation}
\alpha(i\xi_l)=\sum\limits_{j=1}^{10}
\frac{g_j}{\omega_{aj}^2+\xi_l^2},
\label{e20}
\end{equation}
\noindent 
where $g_j$ are the oscillator strengths and $\omega_{aj}$ are
the eigenfrequencies. For the hydrogen atom the values of these
quantities are listed in Table I (note that 
$1\,\mbox{a.u.\ of\  energy}=4.3597\times 10^{-18}\,\mbox{J}=
27.11\,$eV).
Note also that before the substitution in Eqs.~(\ref{e1}) or (\ref{e19})
the atomic dynamic polarizability from Eq.~(\ref{e20}) should be
expressed in cubic meters including the transformation
factor for
$1\,\mbox{a.u.\ of\  polarizability}=1.482\times 10^{-31}\,\mbox{m}^3$.

In addition to the precise representation (\ref{e20}), the atomic
dynamic polarizability of hydrogen atom can be expressed in terms
of a more simple single oscillator model
\begin{equation}
\alpha(i\xi_l)=\frac{g_a}{\omega_a^2+\xi_l^2},
\label{e21}
\end{equation}
\noindent
where $g_a=\alpha_a(0)\omega_a^2$ is expressed through the static
atomic polarizability $\alpha_a(0)=4.50\,$a.u. and the characteristic
energy $\omega_a=11.65\,$eV \cite{34}.

Below we will check that after the substitution to the Lifshitz-type
formulas both expressions (\ref{e20}) and (\ref{e21}) lead to
equal results in the limits of required accuracy. This permits
to use a more simple Eq.~(\ref{e21}) in computations.

It is well known that for hydrogen molecule the single oscillator
model for the dynamic polarizability is more exact than for the
atom. For this reason it is acceptable to present the molecular
dynamic polarizability of hydrogen in the form
\begin{equation}
\alpha(i\xi_l)=\frac{g_m}{\omega_m^2+\xi_l^2},
\label{e22}
\end{equation}
\noindent
where $g_m=\alpha_m(0)\omega_m^2$. Here the static polarizability
and the characteristic energy of hydrogen molecule are equal to
$\alpha_m(0)=5.439\,$a.u. and $\omega_m=14.09\,$eV,
respectively \cite{34}.

Now let us consider the problem of dielectric permittivities of
graphite $\varepsilon_x$ and $\varepsilon_z$ along the imaginary
frequency axis. Both these quantities can be computed with the
help of Kramers-Kronig relation
\begin{equation}
\varepsilon_{x,z}(i\xi)=1+\frac{2}{\pi}
\int_{0}^{\infty}d\omega
\frac{\omega\mbox{Im}\varepsilon_{x,z}(\omega)}{\omega^2+\xi^2}.
\label{e23}
\end{equation}
\noindent
The imaginary parts of the respective dielectric permittivities
along the real axis, in turn, are equal to
$2\mbox{Re}n_{x,z}(\omega)\times\mbox{Im}n_{x,z}(\omega)$, i.e., are
expressed through the real and imaginary parts of the complex
refractive index of graphite for ordinary and extraordinary
rays, respectively.

Ref.~\cite{35} contains the measurement data for 
$\mbox{Re}n_{x,z}(\omega)$ and $\mbox{Im}n_{x,z}(\omega)$
of graphite obtained by different authors in the frequency
region from $\Omega_1=0.02\,$eV to $\Omega_2=40\,$eV
($1\,\mbox{eV}=1.519\times 10^{15}\,$rad/s).
The use of these data to calculate $\varepsilon_{x,z}(i\xi)$
by Eq.~(\ref{e23}) is, however, complicated by the two problems.
First, the interval $[\Omega_1,\Omega_2]$ is too narrow to
calculate $\varepsilon_{x,z}(i\xi)$ at all Matsubara frequencies
contributing to the van der Waals force (by comparison, for Au
the complex refractive index is measured from
0.125\,eV to 10000\,eV). Second, although for $n_x$ data by different
authors are in agreement, in the case of $n_z$ there are contradictory
data in literature at $\omega\leq 15.5\,$eV.

The first problem can be solved by the use of extrapolation.
According to Ref.~\cite{35}, at high frequencies $\omega\geq\Omega_2$
the imaginary parts of graphite dielectric permittivities can be
presented analytically in the form
\begin{equation}
\mbox{Im}\varepsilon_{x,z}^{(h)}(\omega)=
\frac{A_{x,z}}{\omega^3}.
\label{e24}
\end{equation}
\noindent
Here the values of constants
$A_x=9.60\times 10^3\,{\mbox{eV}}^3$ and
$A_z=3.49\times 10^4\,{\mbox{eV}}^3$ 
are determined from the condition of a smooth joining with the
tabulated data at
$\omega=\Omega_2$ \cite{35}.

At low frequencies $\omega\leq\Omega_1$ one may approximate
$\mbox{Im}\varepsilon_x$ with the help of the Drude model \cite{25}
\begin{equation}
\mbox{Im}\varepsilon_{x}^{(l)}(\omega)=
\frac{\omega_p^2\gamma}{\omega(\omega^2+\gamma^2)},
\label{e25}
\end{equation}
\noindent
where the plasma frequency
$\omega_p=1.226\,$eV and the relaxation parameter
$\gamma=0.04\,$eV are determined from the demand of smooth joining
with tabulated data at $\omega=\Omega_1$.

The extrapolation of tabulated data for $\mbox{Im}\varepsilon_z$
to the region of low frequencies is connected with the second 
problem discussed above, i.e., with the contradictory measurements
by different authors. Thus, the measurement data for $n_z(\omega)$
in Ref.~\cite{36} differ considerably from the same data in
Ref.~\cite{37} in the frequency region $\omega\leq 15.5\,$eV.
According to both Refs.~\cite{36,37}, the imaginary part of
$\varepsilon_z(\omega)$ can be extrapolated to low frequencies
$\omega\leq\Omega_1$ by a constant:
\begin{equation}
\mbox{Im}\varepsilon_z^{(l)}(\omega)=
\varepsilon_{z0}^{\prime\prime}=\mbox{const}.
\label{e26}
\end{equation}
\noindent
The values of this constant, however, are found to be different:
$\varepsilon_{z0}^{\prime\prime}=3$ according to Ref.~\cite{37} and
$\varepsilon_{z0}^{\prime\prime}=0$ according to Ref.~\cite{36}.

As a result, the calculation of graphite dielectric permittivities
along the imaginary frequency axis by Eq.~(\ref{e23}) is performed
as follows:
\begin{eqnarray}
&&
\varepsilon_{x,z}(i\xi)=1+\frac{2}{\pi}
\int_{0}^{\Omega_1}d\omega
\frac{\omega\mbox{Im}\varepsilon_{x,z}^{(l)}}{\omega^2+\xi^2}
\nonumber \\
&&\phantom{aaa}
+\frac{2}{\pi}
\int_{\Omega_1}^{\Omega_2}d\omega
\frac{\omega\mbox{Im}\varepsilon_{x,z}^{(t)}}{\omega^2+\xi^2}
+\frac{2}{\pi}
\int_{\Omega_2}^{\infty}d\omega
\frac{\omega\mbox{Im}\varepsilon_{x,z}^{(h)}}{\omega^2+\xi^2},
\label{e27}
\end{eqnarray}
\noindent
where $\mbox{Im}\varepsilon_{x,z}^{(t)}$ 
is found from the tables and
$\mbox{Im}\varepsilon_{x,z}^{(h,l)}$ are given by 
Eqs.~(\ref{e24})--(\ref{e26}). 
Substituting Eqs.~(\ref{e24})--(\ref{e26}) in Eq.~(\ref{e27})
one finds
\begin{eqnarray}
&&
\varepsilon_{x}(i\xi)=1+\frac{2}{\pi}
\frac{\xi\mbox{Arctan}\frac{\Omega_1}{\gamma}-
\gamma\mbox{Arctan}\frac{\Omega_1}{\xi}}{\xi(\xi^2-\gamma^2)}
\omega_p^2
\nonumber \\
&&\phantom{aaa}
+\frac{2}{\pi}
\int_{\Omega_1}^{\Omega_2}d\omega
\frac{\omega\mbox{Im}\varepsilon_{x}^{(t)}(\omega)}{\omega^2+\xi^2}
+\frac{A_x}{\xi^2}\left[\frac{2}{\pi\Omega_2}+\frac{1}{\xi}
\left(\frac{2}{\pi}\mbox{Arctan}\frac{\Omega_2}{\xi}-1\right)
\right],
\nonumber \\
&& \label{e28} \\
&&
\varepsilon_{z}(i\xi)=1+\frac{\varepsilon_{z0}^{\prime\prime}}{\pi}
\ln\left(1+\frac{\Omega_1}{\xi}\right)
+\frac{2}{\pi}
\int_{\Omega_1}^{\Omega_2}d\omega
\frac{\omega\mbox{Im}\varepsilon_{z}^{(t)}(\omega)}{\omega^2+\xi^2}
\nonumber \\
&& \phantom{aaaaaaaa}
+\frac{A_x}{\xi^2}\left[\frac{2}{\pi\Omega_2}+\frac{1}{\xi}
\left(\frac{2}{\pi}\mbox{Arctan}\frac{\Omega_2}{\xi}-1\right)
\right].
\nonumber
\end{eqnarray}

The calculational results from Eq.~(\ref{e28}), obtained by the use of the
tabulated optical data of Refs.~\cite{35,36,37}, are shown in Figs.~2a,b
in the frequency range from $\xi_1=2.47\times 10^{14}\,$rad/s to
$\xi_{2000}$ at $T=300$K. These results allow the precise calculation
of the van der Waals interaction by Eqs.~(\ref{e1}), (\ref{e19}) in
the separation region $a\geq 3\,$nm (note that with the increase of
separation the number of Matsubara frequencies, giving a non-negligible
contribution to the result, decreases). As to the contribution of zero
Matsubara frequency $\xi_0=0$, there is the analytical result
$r_{\|;g}^{s,p}(0,k_{\bot})=1$ which follows from
$\varepsilon_x(i\xi)\to\infty$ when $\xi\to 0$ in accordance with
Eq.~(\ref{e28}). Note that at zero frequency the other reflection
coefficient $r_{\bot;g}^{s,p}(0,k_{\bot})$ does not contribute to the
result due to the multiple $\xi_0^2$ in the right-hand sides of
Eqs.~(\ref{e1}) and (\ref{e19}).

The dependence of $\varepsilon_x(i\xi)$ on $\xi$ in Fig.~2a is typical 
for good conductors (compare with Refs.~\cite{27,28} for Al and Au).
In Fig.~2b the solid line is obtained with the results of Ref.~\cite{37}
(see also  Ref.~\cite{35}) with $\varepsilon_{z0}^{\prime\prime}=3$.
The dashed line in Fig.~2b is obtained by the data of Ref.~\cite{36}
(see also  Ref.~\cite{35}) using $\varepsilon_{z0}^{\prime\prime}=0$.
It is seen that the dashed line differs markedly from the solid
line in the frequency region $\xi<10^{17}\,$rad/s. The respective
differences in the free energy are discussed in the next section.
It is reasonably safe, however, to prefer the solid line in 
Fig.~2b as giving the correct behavior of $\varepsilon_z$ along
the imaginary frequency axis. In fact the difference between the
two lines is due to the absence of absorption bands near the
frequencies of 5\,eV and 11\,eV in the tabulated data of Ref.~\cite{36}
related to $\varepsilon_z$ (note that in the data for
$\varepsilon_x$ there are absorption bands at these frequencies
in both Refs.~\cite{36,37}). This casts doubts on the measurement
data of Ref.~\cite{36} for $\varepsilon_z$ because from the theory
of graphite band structure \cite{38} it follows that the respective
absorption bands must be present simultaneously in both sets of data
for $\varepsilon_x$ and $\varepsilon_z$.

\section{Calculation of the van der Waals interaction between
\protect{\\}
hydrogen atom or molecule and plane surface of graphite}

We consider the hydrogen atom or molecule at a separation $a$ from the 
hexagonal plane surface $(x,y)$
of a graphite semispace of a flat graphite plate of thickness
$d$. Note that the separation distance between the two plane
hexagonal layers in graphite is approximately 0.336\,nm.
All calculations are performed at separations $a\geq 3\,$nm
where one can neglect the atomic structure of graphite and
describe it in terms of dielectric permittivities 
$\varepsilon_x(\omega)$, $\varepsilon_z(\omega)$ as is done in the
Lifshitz theory. Bearing in mind applications at short separations,
it is instructive to present Eq.~(\ref{e1}) in the form of
nonrelativistic van der Waals interaction (see Introduction)
\begin{equation}
F_E^{s,p}(a,T)=-\frac{C_3^{s,p}(a,T)}{a^3},
\label{e29}
\end{equation}
\noindent
where the van der Waals coefficient $C_3^{s,p}$ [for the case of an atom
near a semispace $(s)$ or a plate $(p)$, respectively] is now a
function of both separation and temperature. For the sake of convenience
in numerical computations, we introduce the nondimensional variables
\begin{equation}
y=2aq_l, \qquad \zeta_l=\frac{2a\xi_l}{c}
\equiv\frac{\xi_l}{\omega_c}
\label{e30}
\end{equation}
\noindent
and express the van der Waals coefficient in terms of these variables
\begin{eqnarray}
&&
C_3^{s,p}(a,T)=\frac{k_BT}{8}\left\{2\alpha(0)+
\sum\limits_{l=1}^{\infty}\alpha(i\zeta_l\omega_c)
\right.
\label{e31} \\
&&\phantom{aa}
\times\left.
\int_{\zeta_l}^{\infty}dye^{-y}\left[2y^2
r_{\|;g}^{s,p}(\zeta_l,y)+\zeta_l^2\left[
r_{\bot;g}^{s,p}(\zeta_l,y)-r_{\|;g}^{s,p}(\zeta_l,y)
\right]\right]\right\}.
\nonumber
\end{eqnarray}
\noindent
Note that for separations up to a few hundred nanometers Eq.~(\ref{e31})
practically does not depend on temperature.

In terms of the new variables (\ref{e30}) the coefficients of reflection
from a graphite semispace (\ref{e5}) are rearranged as
\begin{eqnarray}
&&
r_{\|;g}^s(\zeta_l,y)=\frac{\sqrt{\varepsilon_{xl}\varepsilon_{zl}}y-
f_z(y,\zeta_l)}{\sqrt{\varepsilon_{xl}\varepsilon_{zl}}y+
f_z(y,\zeta_l)},
\nonumber \\
&& \label{e32} \\
&&
r_{\bot;g}^s(\zeta_l,y)=\frac{f_x(y,\zeta_l)-y}{f_x(y,\zeta_l)+y},
\nonumber
\end{eqnarray}
\noindent
where
\begin{eqnarray}
&&
f_z^2(y,\zeta_l)=y^2+\zeta_l^2(\varepsilon_{zl}-1), 
\nonumber \\
&&
f_x^2(y,\zeta_l)=y^2+\zeta_l^2(\varepsilon_{xl}-1).
\label{e33}
\end{eqnarray}

In analogy, the reflection coefficients (\ref{e7}) from a flat
plate of thickness $d$ take the form
\begin{eqnarray}
&&
r_{\|;g}^p(\zeta_l,y)=\frac{{\varepsilon_{xl}\varepsilon_{zl}}y^2-
f_z^2(y,\zeta_l)}{{\varepsilon_{xl}\varepsilon_{zl}}y^2+
f_z^2(y,\zeta_l)+2\sqrt{\varepsilon_{xl}\varepsilon_{zl}}y
f_z(y,\zeta_l)\mbox{coth}\left[f_z(y,\zeta_l)d/(2a)\right]},
\nonumber \\
&& \label{e34} \\
&&
r_{\bot;g}^p(\zeta_l,y)=\frac{f_x^2(y,\zeta_l)-y^2}{y^2+f_x^2(y,\zeta_l)+
2yf_x(y,\zeta_l)\mbox{coth}\left[f_x(y,\zeta_l)d/(2a)\right]}.
\nonumber
\end{eqnarray}

Now we substitute the reflection coefficients from a semispace
(\ref{e32}), the precise atomic dynamic polarizability (\ref{e20}) and
data of Fig.~2a for $\varepsilon_x$ and Fig.~2b (solid line)
for $\varepsilon_z$ into Eq.~(\ref{e31}). The calculational results
for the coefficient of van der Waals interaction between a
hydrogen atom and graphite semispace are presented in Fig.~3a by the 
solid line. For comparison the dashed line in Fig.~3a shows the
results obtained with the use of alternative data for $\varepsilon_z$
(dashed line in Fig.~2b). As is seen from Fig.~3a, at the shortest
separation $a=3\,$nm the use of the alternative data for $\varepsilon_z$
leads to a 15\% error in the value of the van der Waals coefficient
which decreases with an increase of separation.

The computation of $C_3^s$ was repeated using the single oscillator model
(\ref{e21}) for the atomic dynamic polarizability instead of the
10-oscillator model (\ref{e20}). The results were found to be
practically in coincidence with those in Fig.~3a (the maximum deviations 
are less than 0.2\% in the separation region from 3\,nm to 150\,nm). Thus,
the single oscillator model is a sufficient approximation for the
atomic (and, consequently, molecular) dynamic polarizability of
hydrogen in computations of the short-range van der Waals interaction
with a graphite surface.

In the same way as above, we calculate the van der Waals coefficient
$C_3^s$ for the interaction of a hydrogen molecule with graphite
semispace. The only difference is the use of the molecule dynamic
polarizability (\ref{e22}) instead of atomic one. The results are
shown in Fig.~3b by the solid line (the dashed line is calculated
by the less accurate alternative data of Ref.~\cite{36} for the 
dielectric permittivity $\varepsilon_z$). The comparison of
Figs.~3a and 3b leads to the conclusion that the magnitudes of the
van der Waals coefficient for the hydrogen molecule are larger than
for the atom.

Now let the hydrogen atom be located at a separation $a$ from the
flat graphite plate of thickness $d$. Of interest is the dependence
on $d$ of the van der Waals free energy of atom-plate interaction.
The calculations of the free energy were performed by Eqs.~(\ref{e29})
and (\ref{e31}) with reflection coefficients (\ref{e32}) (for a
semispace) and (\ref{e34}) (for a plate of thickness $d$). The values
of dielectric permittivities along the imaginary frequency axis were
taken from Fig.~2 (solid lines) and the atomic dynamic polarizability
from Eq.~(\ref{e21}). In Fig.~4 the ratios of the free energies are
plotted for the case of a plate and a semispace as a function of
plate thickness for hydrogen atom located at different separations
from the graphite surface (line 1 for $a=3\,$nm, line 2 for $a=10\,$nm,
line 3 for $a=20\,$nm, and line 4 for $a=50\,$nm). As is seen from
Fig.~4, at a separation $a=3\,$nm the finite thickness of the plate
has a pronounced effect on the free energy (more than 1\% change)
only for thcknesses $d<8\,$nm. At separations $a=10\,$nm, 20\,nm
and 50\,nm the finite thickness of the plate leads to a smaller
magnitude of the van der Waals free energy, as compared with a
semispace, for more than 1\% if the thickness of a plate is less
than 19\,nm, 32\,nm and 61\,nm, respectively. Thus, if the
separation between an atom and a plate is $a=3\,$nm, then the plate
of $d=8\,$nm thickness can be already considered with a good
accuracy as a semispace.

\section{Calculation of the van der Waals interaction for
\protect{\\}
hydrogen atom or molecule external to multiwall carbon nanotube}

The multiwall carbon nanotube can be modelled by a graphite
cylindrical shell of some length $L$, external radius $R\ll L$ and 
thickness $d<R$. In doing so  the hexagonal  layers of graphite crystal
lattice form the external surface of a cylinder and the internal
sections concentric to it. The crystal optical axis $z$ is
perpendicular to the surface of the cylinder at each point.
The above derived Lifshitz-type formula (\ref{e19}) is applicable
to the case of multiwall carbon nanotube if its thickness $d$ is
large enough (typically $D\geq 3\,$nm), so that the nanotube contains
sufficiently many layers. Then it is possible to neglect the
atomic structure of graphite and to describe it in terms of
dielectric permittivity.

For convenience in numerical computations we rewrite Eq.~(\ref{e19})
in terms of dimensionless variables (\ref{e30}) representing the free
energy of the van der Waals interaction with a cylinder in the form
\begin{equation}
F_E^c(a,T)=-\frac{C_3^c(a,T)}{a^3},
\label{e35}
\end{equation}
\noindent
where
\begin{eqnarray}
&&
C_3^{c}(a,T)=\frac{k_BT}{8}
\sqrt{\frac{R}{R+a}}\left\{
\vphantom{\sum\limits_{l=1}^{\infty}\int\limits_{\zeta_l}^{\infty}}
\frac{4R+3a}{2(R+a)}\alpha(0)\right.
\nonumber \\
&&\phantom{aaa}
+\sum\limits_{l=1}^{\infty}
\alpha(i\zeta_l\omega_c)
\int_{\zeta_l}^{\infty}dy ye^{-y}
\left[y-\frac{a}{2(R+a)}\right]
\nonumber \\
&&\phantom{aaa}
\times
\left.\left[
2r_{\|;g}^{s,p}(\zeta_l,y)+
\frac{\zeta_l^2}{y^2}\left[
r_{\bot;g}^{s,p}(\zeta_l,y)-
r_{\|;g}^{s,p}(\zeta_l,y)\right]\right]\right\}.
\label{e36}
\end{eqnarray}
\noindent
The reflection coefficients were defined in Eq.~(\ref{e32}) (with
index $s$ related to the case of a solid cylinder) and in 
Eq.~(\ref{e34}) (with index $p$ related to the case of a cylindrical
shell of thickness $d$).

Let us first compare the van der Waals interaction between hydrogen 
atom or molecule with a graphite semispace and a solid cylinder.
The differences of the interaction strength with a semispace and
a cylinder can be characterized by a parameter
$\delta=(C_3^s-C_3^c)/C_3^s$. A few results for a graphite cylinder
with $R=50\,$nm, calculated by Eqs.~(\ref{e36}), (\ref{e31}),
(\ref{e21}), (\ref{e22}) and dielectric permittivities given by the
solid lines of Fig.~2, are presented in Table II (columns 2--4 and
5--7 are related to the cases of hydrogen atom and molecule,
respectively). As is seen from Table II, at short separations
of about a few nanometers there are only minor differences between
$C_3^s$ and $C_3^c$. With increase of $a$, however, the magnitude
of $\delta$ quickly increases. This takes place for both hydrogen
atom and molecule.

It is interesting to follow the dependence of the van der Waals
coefficient $C_3^c$ on $R$ for atoms and molecules located at different
separations from the cylinder surface. These computations were performed 
with Eqs.~(\ref{e36}), (\ref{e5}), (\ref{e21}), (\ref{e22}) 
and the same data for graphite dielectric permittivities.
The results are presented in Fig.~5a (for hydrogen atom) and Fig.~5b
(for hydrogen molecule) where the lines 1, 2 and 3 are pictured
for separations $a=3\,$nm, 5\,nm and 10\,nm, respectively.
It is seen that with the increase of $R$ the van der Waals coefficients
are also increasing.

Now consider the cylindrical shell of radius $R$ and thickness $d$
with the longitudinal cavity of a radius
$R-d$. This is evidently a better model for a multiwall carbon
nanotube. In Fig.~6 we present the computation results for the interaction
between a hydrogen atom and a cylindrical envelope with $R=20\,$nm as
a function of envelope thickness $d$ (atom is located at a separation
$a=5\,$nm from the external surface of the cylindrical shell). 
The computations
were performed by Eq.~(\ref{e36}) using the same procedure as above. 
The value $d=20\,$nm corresponds to the case of a solid cylinder.
It is interesting, however, that already at $d=11\,$nm the magnitude
of $C_3^c$ is only 1\% lower than the one obtained for the solid
cylinder of $R=20\,$nm radius. For less thickness of the cylindrical
shell the smaller values of the van der Waals coefficient are
obtained (the same is true also for a hydrogen molecule). Note that
we do not extend the line of Fig.~6 for thicknesses less than 3\,nm
where the macroscopic description of graphite in terms of dielectric
permittivity may be not applicable.

\section{Comparison between the free energies of
hydrogen atoms inside and outside of multiwall carbon nanotubes}

The obtained above Lifshitz-type formulas (\ref{e19}), (\ref{e36})
provides a good approximate description of the van der Waals interaction
when a microparticle is located outside of a cylindrical shell.
Let us now consider a microparticle inside of the same shell.
In this case the van der Waals free energy can be approximately
calculated by the method of pairwise summation of the interatomic
potentials with subsequent normalization of the obtained interaction
coefficient using the known case of microparticle near a semispace
\cite{12,39}. For a microparticle outside of an arbitrary macrobody
$v$ this method leads to the expression
\begin{equation}
F_E^v(a,T)\approx-\frac{6C_3^s(a,T)}{\pi}
\int_{v}\frac{dv}{r^6},
\label{e37}
\end{equation}
\noindent
where $r$ is the separation between the microparticle and an atom (molecule) 
of the macrobody.

To determine the accuracy of Eq.~(\ref{e37}), let us apply it in the
case of hydrogen atom outside of a solid graphite cylinder at a separation 
$a$ [to which Eq.~(\ref{e36}) is also applicable]. Then Eq.~(\ref{e37})
is rewritten as
\begin{equation}
F_E^{c,ext}\equiv F_E^c(a,T)\approx -\frac{24C_3^s(a,T)}{\pi}
\int_{0}^{\theta_m}d\theta\int_{0}^{\infty}dz
\int_{\rho_1(\theta)}^{\rho_2(\theta)}
\frac{\rho d\rho}{(\rho^2+z^2)^3},
\label{e38}
\end{equation}
\noindent
where
$\sin\theta_m=R/(R+a)$, $R$ is the cylinder radius, and
$\rho_{1,2}(\theta)$ are the two solutions of the equation
\begin{equation}
\rho^2+(R+a)^2-2\rho (R+a)\cos\theta=R^2.
\label{e39}
\end{equation}
\noindent
After the integration over $z$ and $\rho$ Eq.~(\ref{e38}) takes
the form
\begin{equation}
F_E^{c,ext}(a,T)\approx -\frac{3}{2}C_3^s(a,T)
\int_{0}^{\theta_m}d\theta
\left[\frac{1}{\rho_1^3(\theta)}-
\frac{1}{\rho_2^3(\theta)}\right].
\label{e40}
\end{equation}
\noindent
The numerical computations by Eq.~(\ref{e40}) demonstrate that for
a cylinder with $R=50\,$nm the results, obtained by the method of
additive summation, differ by less than 1\% from the results,
obtained by the Lifshitz-type Eq.~(\ref{e35}), within the separation
range $a\leq 8\,$nm. At $a=10\,$nm the free energies computed by
the two formulas differ for 1.35\%, and at $a=50\,$nm by 16\%.
Hence the method of additive summation works well at small separations
between an atom and a cylindrical surface. This makes it reasonable to
apply this method for hydrogen atom inside of a multiwall carbon
nanotube.

We consider a hydrogen atom inside of a nanotube with thickness $d$
and internal radius $R_0=R-d$ at a separation $a$ from the internal
surface. In accordance with Eq.~(\ref{e37}), the free energy of the
van der Waals interaction is
\begin{equation}
F_E^{c,int}(a,T)\approx -\frac{24C_3^s(a,T)}{\pi}
\int_{0}^{\pi}d\theta\int_{0}^{\infty}dz
\int_{\tilde{\rho}_1(\theta)}^{\tilde{\rho}_2(\theta)}
\frac{\rho d\rho}{(\rho^2+z^2)^3},
\label{e41}
\end{equation}
\noindent
where the integration limits are given by
\begin{eqnarray}
&&
\tilde{\rho}_1(\theta)=-(R_0-a)\cos\theta+
\sqrt{R_0^2-(R_0-a)^2\sin^2\theta},
\nonumber \\
&&
\tilde{\rho}_2(\theta)=-(R_0-a)\cos\theta+
\sqrt{(R_0+d)^2-(R_0-a)^2\sin^2\theta}.
\label{e42}
\end{eqnarray}

After the integration over $z$ and $\rho$ Eq.~(\ref{e41}) leads to
\begin{equation}
F_E^{c,int}(a,T)\approx -\frac{3}{2}C_3^s(a,T)
\int_{0}^{\pi}d\theta
\left[\frac{1}{\tilde{\rho}_1^3(\theta)}-
\frac{1}{\tilde{\rho}_2^3(\theta)}\right].
\label{e43}
\end{equation}

In Fig.~7 we present the results of numerical computations by 
Eq.~(\ref{e43}) for the hydrogen atom inside of the hypothetical
nanotube with the internal radius $R_0=10\,$nm and external
radius $R=50\,$nm. The free energy of the atom-nanotube interaction
is plotted in Fig.~7 as a function of atom position between the
opposite points of the internal cylindrical surface.
The atom positions closer than 3\,nm to the internal surface are
not reflected in the figure (their consideration would demand a
more exact treatment of the atomic structure of graphite).
As is seen from Fig.~7, the free energy reaches a maximum on the 
cylinder axis, where the van der Waals force acting on an atom is 
equal to zero in accordance with symmetry considerations.
This equilibrium state is, however, unstable and under the influence
of fluctuations the hydrogen atom will move to positions with
lower free energy near the internal cylindrical surface of a nanotube.

Now we are in a position to compare the free energies of hydrogen atoms
located outside and inside a multiwall carbon nanotube in order to
decide which position is preferable energetically. In Fig.~8 the
calculation results for the differences of free energies
$F_E^{c,ext}$ and $F_E^{c,int}$ are presented as a function of
thickness of the nanotube. In doing so we consider both atoms,
internal and external, situated at a separation $a=3\,$nm
from the internal and external surfaces of a nanotube, respectively.
The solid line in Fig.~8 is related to the fixed internal radius
of the nanotube $R_0=10\,$nm, and in this case the external radius 
increases together with thickness of the nanotube $d$. The dashed
line is for a fixed external radius $R=50\,$nm and decreasing
internal radius with the increase of $d$. The computations were
performed with Eq.~(\ref{e43}) for a position of the atom inside the 
nanotube and with Eq.~(\ref{e35}) for position of the atom outside
the nanotube.

As is seen from Fig.~8, in all cases the difference between the
external and internal free energies of the van der Waals interaction
is positive. What this means is the position of a hydrogen atom
inside a multiwall carbon nanotube is preferable energetically.
Comparing the solid and dashed lines in Fig.~8, we conclude that
for nanotubes of fixed thickness $d$ the potential well for the
hydrogen atom inside a nanotube is deeper if nanotube has a smaller
external radius $R$. This is an encouraging result which points
to the possibility
of hydrogen storage inside carbon nanostructures.

\section{Conclusions and discussion}

In the above we have widened the scope of the Lifshitz theory of
the van der Waals force by considering new configurations of much
interest which have not been explored previously. The first to be
investigated was the van der Waals force between an atom or molecule 
and a plane surface of an uniaxial crystal perpendicular to the crystal
optic axis. For this configuration the exact expression for the 
free energy of the van der Waals and Casimir-Polder interaction is
given by Eq.~(\ref{e1}) with the reflection coefficients (\ref{e5})
(for the case of a microparticle near a semispace) or (\ref{e7})
(for a microparticle near a plate of finite thickness). We next
derive the approximate Lifshitz-type formula (\ref{e19}) for the 
free energy of the van der Waals interaction between microparticle
and solid cylinder or cylindrical shell having a longitudinal concentric
cavity. This cylinder may be made of isotropic material or of an
uniaxial crystal. The accuracy of the obtained formula was shown to
be of about 1\% at microparticle-cylinder separations less than one half 
of a cylinder radius.

The above extensions of the Lifshitz formula for microparticle-wall
interaction were applied to the case of hydrogen atom or molecule
near a graphite surface. For this purpose the dielectric permittivities
of graphite along the imaginary frequency axis were found by the use
of tabulated optical data for the complex refractive index. In doing so
different sets of data were analyzed and necessary extrapolations to
high and low frequencies were done. Together with the use of hydrogen
atomic and molecular dynamic polarizabilities, this allowed us to
calculate the van der Waals interaction between hydrogen atom or
molecule and graphite semispace, graphite flat plate of finite
thickness or solid graphite cylinder and cylindrical shell. In particular, 
the influence of the thickness of the plate on the van der Waals
interaction was investigated.

The calculation results for the atom-cylinder case were used to model
the van der Waals interaction between hydrogen atoms or molecules
and multiwall carbon nanotube with sufficiently large number of
layers. In particular, the dependence of the van der Waals
interaction of the atom-nanotube case on nanotube thickness was investigated.
Notice that the developed formalism is not applicable to single-
or twowall nanotubes where the atomic structure of the wall should be
taken into account. In this case the van der Waals force can be computed
in the framework of density functional theory \cite{39a,39b,39c}.
 
Finally, we have compared the free energies of the van der Waals
interaction between a hydrogen atom and multiwall carbon nanotube for
the cases when atom is located outside or inside of the nanotube.
It was shown that atoms situated inside of a multiwall nanotube
possess lower free energy in a wide region of nanotube thicknesses,
i.e., such a position is energetically preferable. This conclusion
is promising for the possibility of using carbon nanotubes for the
purpose of hydrogen storage.

Many other opportunities for application of the obtained generalizations 
of the Lifshitz formula in physics of dispersion forces are possible.

\section*{Acknowledgments}
The authors are grateful to J.~F.~Babb for 
stimulating discussions and useful references on the atomic
dynamic polarizability of hydrogen. G.L.K. and V.M.M. were partially
supported by Finep (Brazil).

\begin{figure*}
\vspace*{-5cm}
\includegraphics{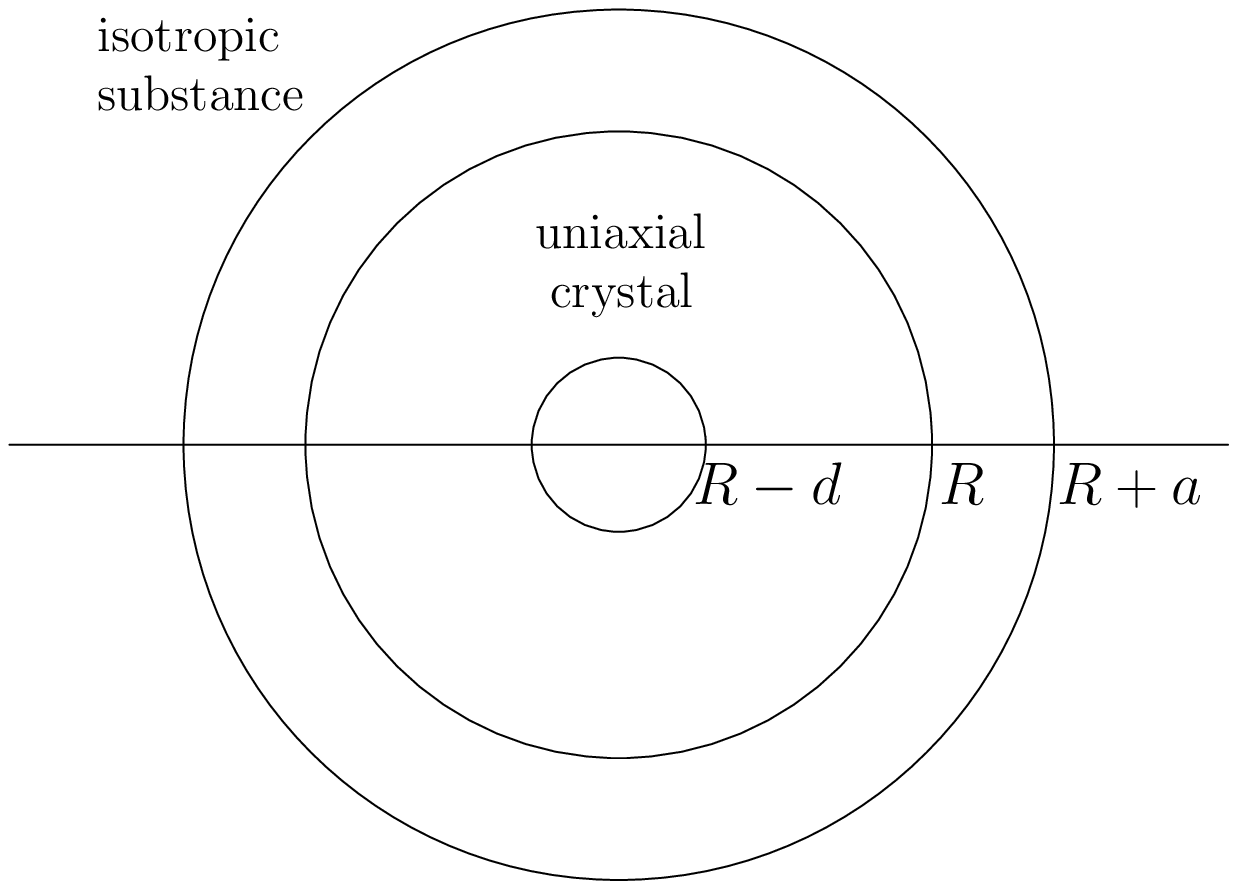}
\vspace*{-12.5cm}
\caption{
Schematic of the cylinder of radius $R$ made of a
 uniaxial crystal and having a longitudinal concentric cavity
of radius $R-d$. This cylinder is concentrically placed into
a cylindrical cavity of radius $R+a$ in the infinite space filled 
with an isotropic substance.
}
\end{figure*}
\begin{figure*}
\vspace*{-2cm}
\includegraphics{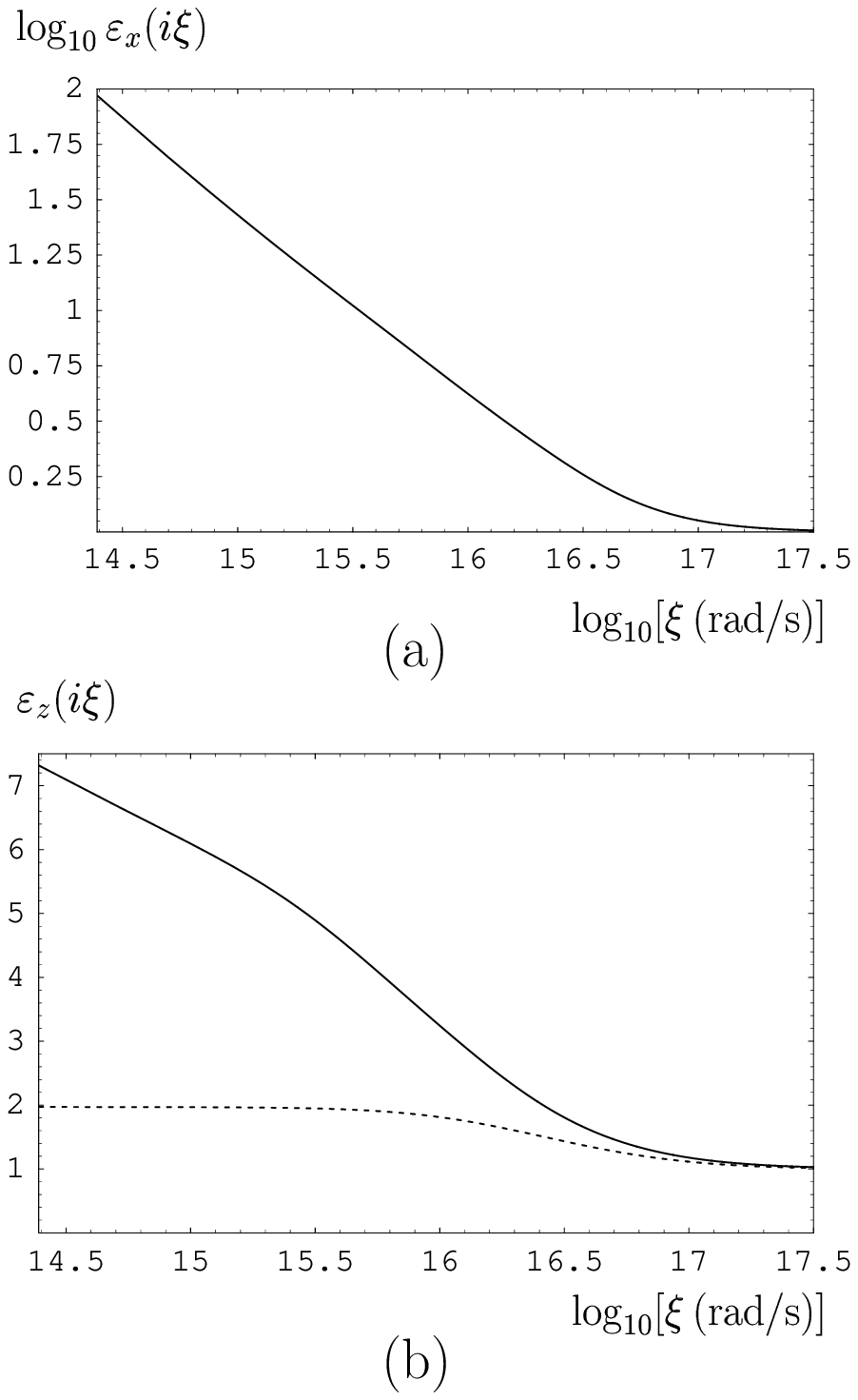}
\vspace*{-11.5cm}
\caption{
Dielectric permittivity of graphite along the imaginary frequency
axis in (a) the hexagonal layer and (b) perpendicular to it, as 
a function of frequency. Solid and dashed lines in (b) are obtained
with the optical data of Ref.~\cite{37} and 
Ref.~\cite{36}, respectively.
}
\end{figure*}
\begin{figure*}
\vspace*{-2cm}
\includegraphics{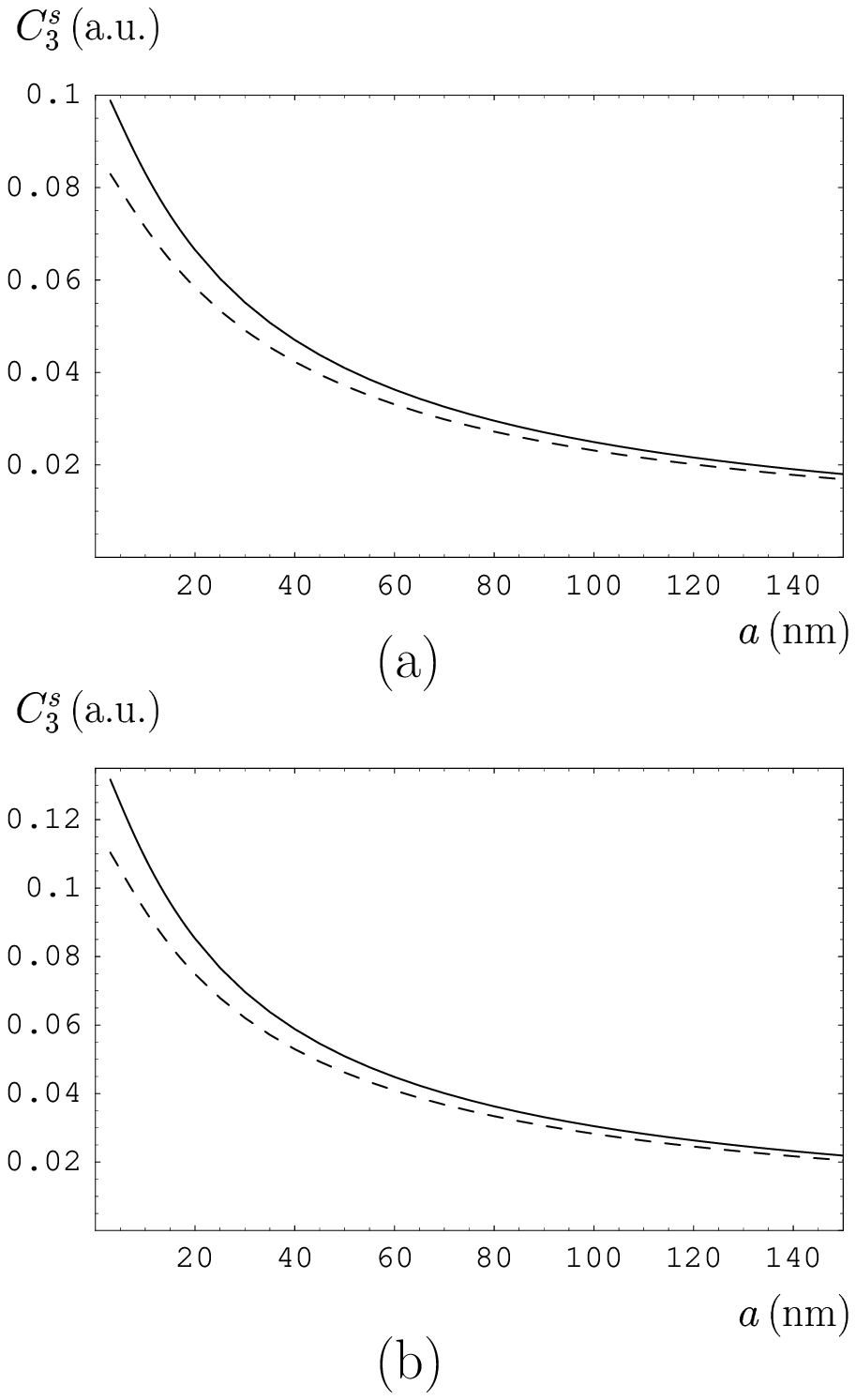}
\vspace*{-11.5cm}
\caption{
Dependence of the van der Waals coefficient $C_3^s$ on separation
of (a) hydrogen atom and (b) molecule, from graphite semispace.
The solid and dashed lines are obtained
with the optical data of Ref.~\cite{37} and 
Ref.~\cite{36}, respectively.
}
\end{figure*}
\begin{figure*}
\vspace*{-7cm}
\includegraphics{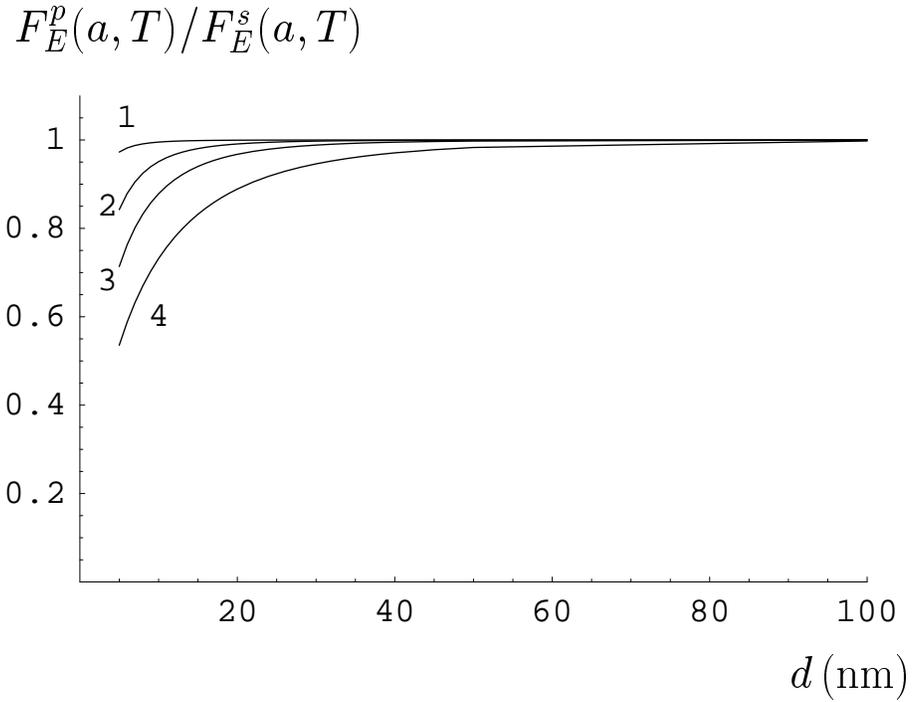}
\vspace*{-8.5cm}
\caption{
The ratios of the free energies for the van der Waals atom-plate to 
atom-semispace interaction as a function of plate thickness
for hydrogen atom located at different separations
from the graphite surface (lines 1, 2, 3 and 4 are for separations
$a=3\,$nm, $10\,$nm, $20\,$nm and $50\,$nm, respectively).
}
\end{figure*}
\begin{figure*}
\vspace*{-2cm}
\includegraphics{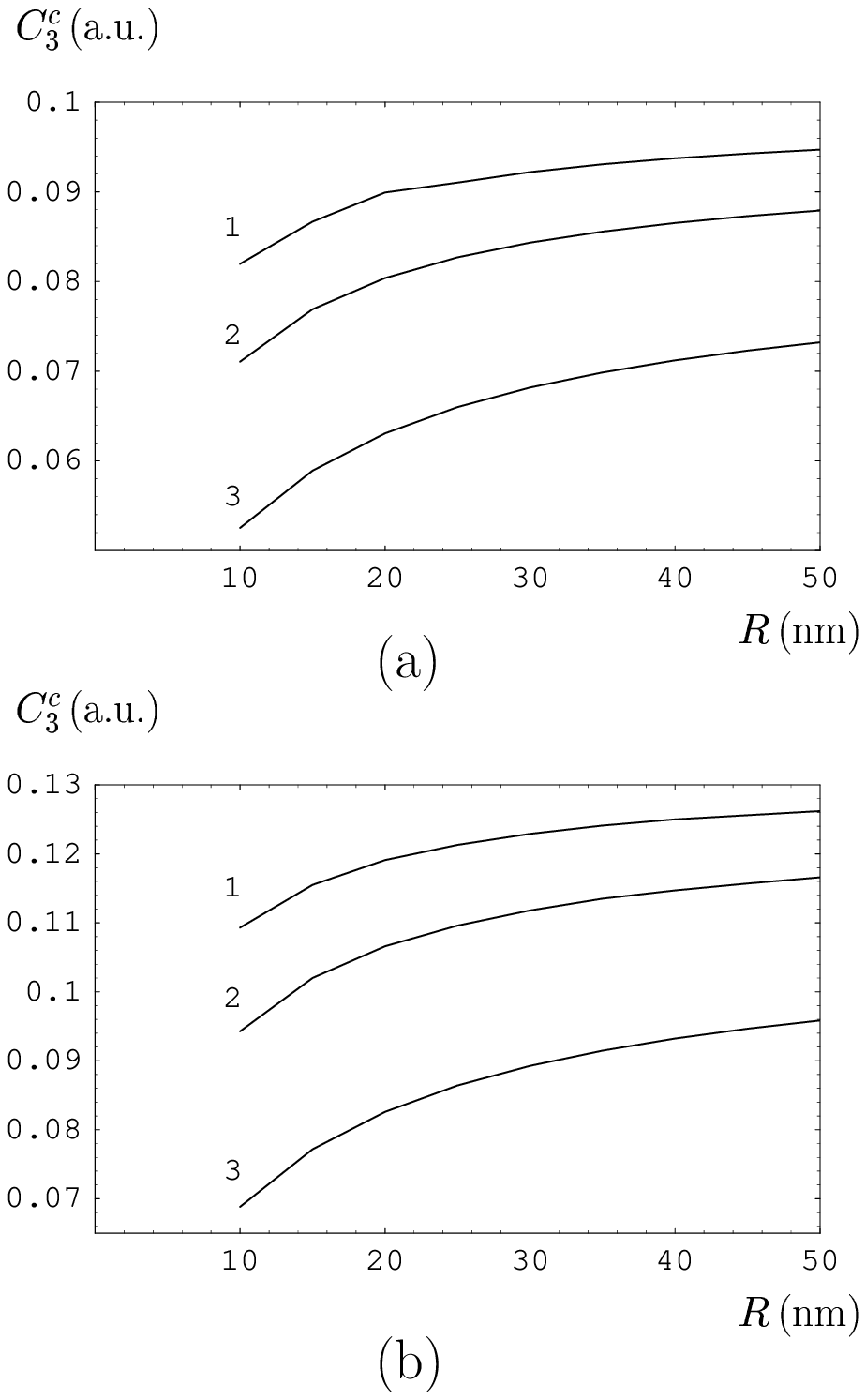}
\vspace*{-11.5cm}
\caption{Dependence of the van der Waals coefficient $C_3^s$ on
the cylinder radius for (a) hydrogen atom and (b) molecule, located
at different separations from the graphite cylinder
(lines 1, 2  and 3 are for separations $a=3\,$nm, $5\,$nm and 
$10\,$nm, respectively).
}
\end{figure*}
\begin{figure*}
\vspace*{-7cm}
\includegraphics{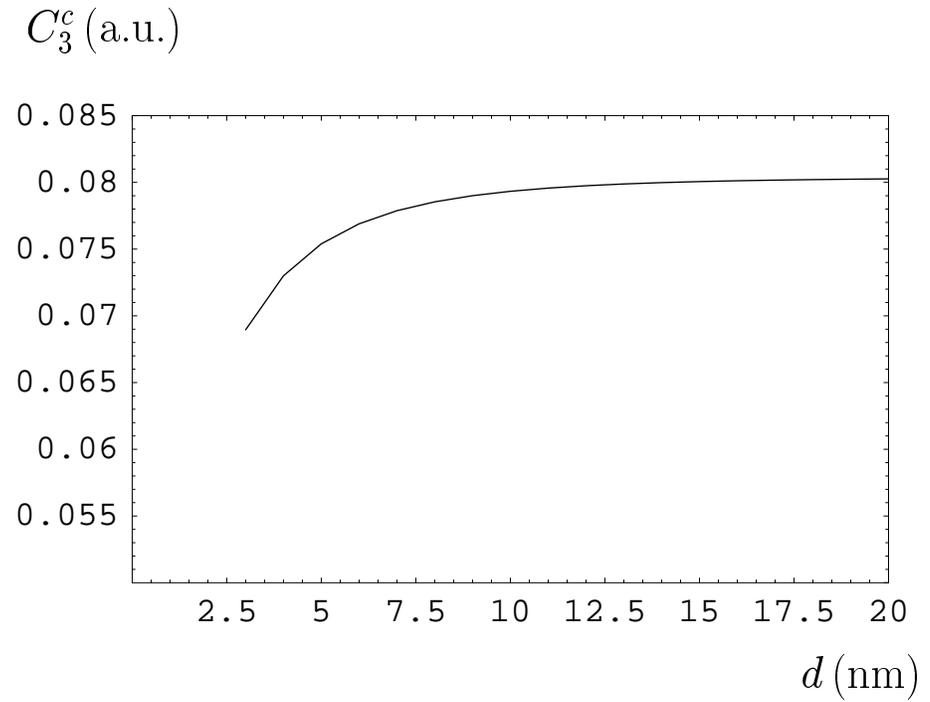}
\vspace*{-11.5cm}
\caption{Dependence of the van der Waals coefficient $C_3^s$ on
thickness of the cylindrical shell with an external radius
$R=20\,$nm for hydrogen atom at a separation $a=5\,$nm from
the shell.
}
\end{figure*}
\begin{figure*}
\vspace*{-7cm}
\includegraphics{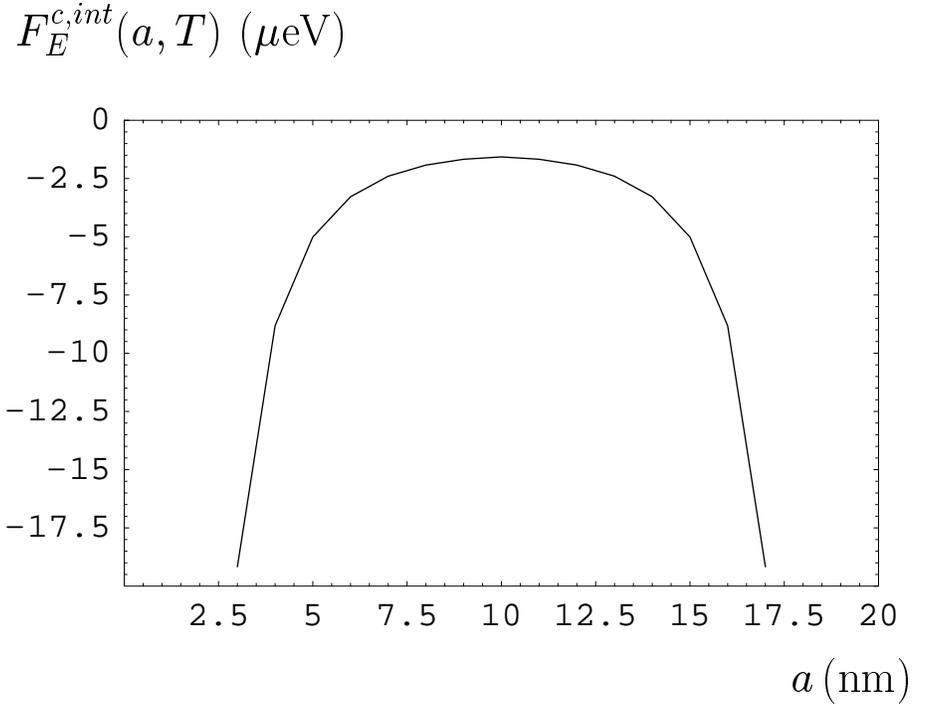}
\vspace*{-11.5cm}
\caption{
The van der Waals free energy for hydrogen atom inside of the carbon
nanotube with internal radius $R_0=10\,$nm and external radius
$R=50\,$nm as a function of the atom position between the opposite
points of the internal cylindrical surface.
}
\end{figure*}
\begin{figure*}
\vspace*{-7cm}
\includegraphics{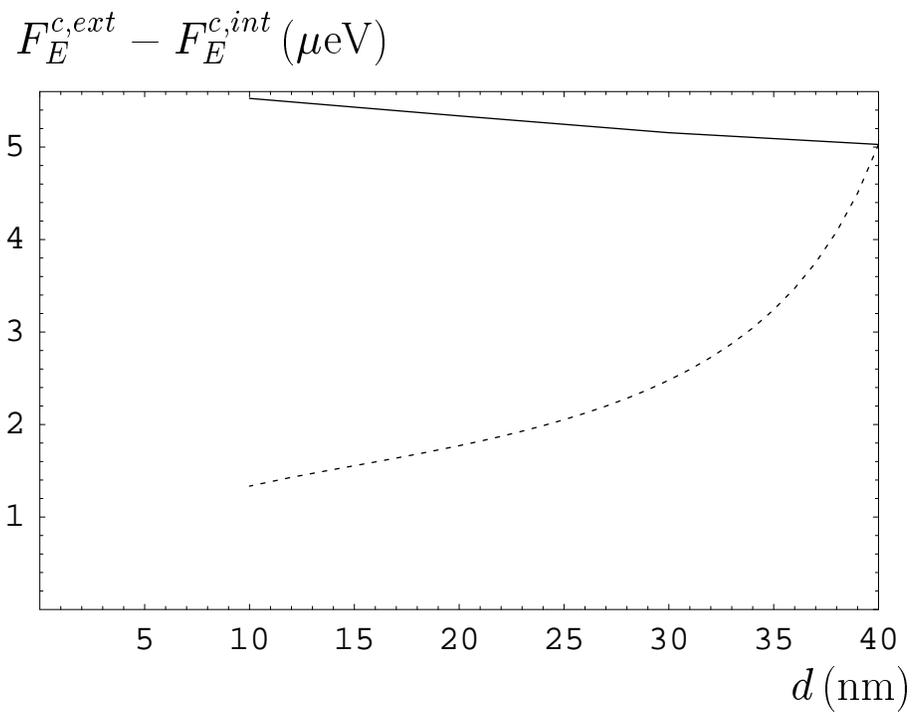}
\vspace*{-11.5cm}
\caption{
Difference of the free energies of hydrogen atoms situated outside 
and inside of the multiwall carbon nanotube as a function of nanotube
thickness. The solid and dashed lines are for the nanotubes with
a fixed internal radius $R_0=10\,$nm and fixed external radius
$R=50\,$nm, respectively.
}
\end{figure*}
\begingroup
\squeezetable
\begin{table}
\caption{The values of strengths and eigenenergies
of oscillators for hydrogen atom in the framework of
the 10-oscillator model.}
\begin{ruledtabular}
\begin{tabular}{|r|l|l|} 
$j$ & $g_j$ & $\omega_{aj}\,$(a.e.) \\
\hline
1 & 0.41619993 & 0.37500006 \\
2 & 0.08803654 & 0.44533064 \\
3 & 0.08993244 & 0.48877611 \\
4 & 0.10723836 & 0.56134416 \\
5 & 0.10489786 & 0.68364018 \\
6 & 0.08700329 & 0.89169023 \\
7 & 0.06013601 & 1.2698693 \\
8 & 0.03259492 & 2.0478339 \\
9 & 0.01199044 & 4.0423429 \\
10 & 0.00197021 & 12.194172
\end{tabular}
\end{ruledtabular}
\end{table}
\endgroup
\begingroup
\squeezetable
\begin{table}
\caption{Magnitudes of the van der Waals coefficients
$C_3^s$ and $C_3^c$ and their relative differences $\delta$
(see text) for the interaction of hydrogen atom or molecule
with a graphite semispace or a cylinder with radius
$R=50\,$nm.}
\begin{ruledtabular}
\begin{tabular}{|r|l|l|r|l|l|r|} 
\multicolumn{1}{|c|}{$a$}&
\multicolumn{3}{c|}{$H$}&
\multicolumn{3}{c|}{$H_2$} \\ \hline
(nm)& $C_3^s\,$(a.u.) & $C_3^c\,$(a.u.) & $\delta$(\%) &
$C_3^s\,$(a.u.) & $C_3^c\,$(a.u.) & $\delta$(\%) \\ \hline
3 & 0.09882 & 0.09471 & 4.2 &0.1317 &0.1262 &4.2 \\
5 & 0.09416 &0.08792 & 6.6 & 0.1248 &0.1166 & 6.6 \\
10 & 0.08316 & 0.07322 & 12.0 & 0.1088 &0.09584 & 11.9 \\
20 &0.06652 & 0.05301 & 20.3 & 0.08526 & 0.06801 & 20.2 \\
30 & 0.05516 & 0.04047 & 26.6 & 0.06970 & 0.05118 & 26.6 \\
40 & 0.04704 & 0.03214 & 31.7 & 0.05885 & 0.04025 & 31.6 \\
50 & 0.04098 & 0.02631 & 35.8 & 0.05090 & 0.03270 & 35.8 
\end{tabular}
\end{ruledtabular}
\end{table}
\endgroup
\end{document}